\documentclass[preprintnumbers,showpacs,amsmath,amssymb,floatfix,9pt,prd,twocolumn,
superscriptaddress,nofootinbib]{revtex4}

\usepackage{latexsym}
\usepackage{epsfig}
\usepackage{amssymb}
\usepackage{float}
\usepackage{color}
\usepackage{hyperref}

\begin{document}

\title{Stability of a $d$-dimensional thin-shell wormhole surrounded by quintessence}

\author{Ayan Banerjee}
\email{ayan_7575@yahoo.co.in} \affiliation{Department of
Mathematics, Jadavpur University, Kolkata 700 032, West Bengal,
India}
\author{Kimet Jusufi}
\email{kimet.jusufi@unite.edu.mk} \affiliation{Physics Department, State University of Tetovo, Ilinden Street nn, 1200, Tetovo, Macedonia}
\author{Sebastian Bahamonde}
\email{sebastian.beltran.14@ucl.ac.uk}\affiliation{Department of Mathematics, University College London,
	Gower Street, London, WC1E 6BT, UK}

\begin{abstract}
We study the stability of different higher dimensional thin--shell wormholes (HDTSW) in general relativity with a cosmological constant. We show that a $d$--dimensional thin--shell wormhole surrounded by quintessence can have three different throat geometries: spherical, planar and hyperbolic. Unlike the spherical geometry, the planar and hyperbolic geometries allow different topologies that can be interpreted as higher-dimensional domain
walls or branes connecting two universes. To construct these geometries, we use the cut-and-paste procedure
by joining two identical vacuum spacetime solutions. Properties such as the null energy
condition and geodesics are also studied. A linear stability analysis around the static solutions is carried
out. Our stability analysis takes into account a more general HDTSW geometry than previous works so  it is possible to recover other well-known stability HDTSW conditions. 
\end{abstract}


\maketitle

\section{Introduction}
One of the most interesting solutions in General Relativity (GR) are known as wormhole solutions. GR predicts that the spacetime is deformable due to energy/matter allowing the existence of this exotic geometries. Basically, a wormhole is a tunnel which connects two different asymptotically flat regions from the spacetime. The first ideas related to wormhole geometries were suggested by Flamm \cite{LF} and later by Einstein and Rosen \cite{EN}. The later introduced the so-called ``Einstein-Rosen bridge" constructed by the Schwarzschild solution by using the idea of two black holes connecting two different regions of spacetime. However, it was demonstrated that this geometry cannot be traversable due to the singularity. In 1962, Fuller and Wheeler \cite{RJ}
employed Kruskal coordinates to describe the geometry of the Schwarzschild wormhole showing that it is not traversable. They showed that if that wormhole is opened, it would close so quickly that even a single photon could not be able to travel through it. The interest of studying wormholes geometries was highly stimulated
after the work of traversable wormholes done by Morris \& Thorne \cite{mo}. They constructed transversable wormhole geometries as tunnels from one region of spacetime to another. They showed that in order to travel through this region, the  matter supporting those geometries necessarily need to violate the standard energy conditions \cite{mv,mo,jps,mm}. Hence, in the framework of GR, the so-called ``exotic matter" which violates the standard energy conditions is needed to have a transversable wormhole. Since this kind of matter has not been directly observed in the nature, it is a problematic theoretical issue to deal. One possible way to study this problem is by introducing thin-shell wormholes. One can restrict the violation of the energy conditions to
an infinitesimally small thin-shell allowing to construct wormhole which violates the energy conditions only in this region \cite{mvi,mvis}. Therefore, even if we can not ignore it completely, one can concentrate the study on a thin-shell. Thus, using the cut-and-paste
technique, one can concentrate the exotic matter at the wormhole throat.
The surface stress-energy tensor components at the throat are determined
invoking the Darmois-Israel formalism \cite{iw,pa}, which leads to the Lanczos equations \cite{na,kl,pm}. If the equation of state for the matter on the shell is provided, we can obtain the dynamical
evolution of the wormhole  using the solution of the Lanczos equation.

 In this scenario, stability analysis of wormholes and thin-shell wormholes under linear perturbations preserving the
original symmetries have been carried out by several authors. Poisson and Visser \cite{po} performed a
stability analysis for the Schwarzschild thin-shell wormhole without assuming specific equations
of state (EoS) for the exotic matter. Considering the same stability analysis, more general geometries have been studied in the literature such as incorporating a charge, a cosmological constant or another physical quantities (see, for example,
references \cite{efe,fsn,efc,fr,mt,mg,gd,aa,frk,mgr,jpsle,jpsl,kab}). Stability analysis of
cylindrical thin-shell wormholes have been studied in \cite{Sharif,SHM,MRS}.
Different thin-shell wormholes with different EoS have been also discussed in various works \cite{MPF,CE,E,FSN,S,TME,P}.

In 1920, Kaluza and Klein unified electromagnetism
	and gravity by introducing a 5 dimensional theory. According to the Kaluza-Klein picture, extra dimensions are compactified to a small radius at the order of the Planck length when one is considering low energy physics.	String theory/M-Theory predicts the existence of extra dimensions at low energies and introduced the notion of Braneworlds. These theories are promising candidates for
	a quantum theory of gravity and then for the unification of  gravity and quantum effects. This motivates the study of geometries in higher dimensional spacetimes.
In Braneworld scenarios \cite{nah,ia,lr,lra}, the Standard Model of particles
is confined on a 3-brane, where only gravity is allowed to penetrate
the extra dimensions. In this context, the 4-dimensional universe is viewed as a hypersurface
called the ``brane", which is embedded in a higher dimensional spacetime called  ``the bulk". Under these models, all the matter fields are confined to the Brane with a variety of different mechanisms, while gravity can propagate in the bulk. During the last two decades, to understand in a more general way how gravity behaves in higher dimensions, various
important solutions of the Einstein equations in higher dimensions have been studied (for example see \cite{rc,vn,uk,ka,sep,hk,jp})

The study of GR in higher dimensions gives the additional motivation to find possible factors and interesting features that are not revealed in dimensions other	than four.  Hence, the study of higher dimensional wormholes or thin-shell wormholes are also a very interesting subject to analyse. There are different interesting studies related to Lorentzian and also Euclidean wormholes in higher dimensional gravity. Euclidean wormholes have been studied by Jianjun and	Sicong and by Gonzales-Diaz \cite{pg,xj}. Self-dual Lorentzian wormholes have been also studied in
the context of $N$-dimensional Einstein gravity by Mauricio Cataldo et al.\cite{mc}. They have
also studied $(N+1)$-dimensional evolving wormholes supported by matter satisfying a polytropic
equation of state \cite{mf}. Other higher dimensional evolving wormholes have been studied, see for example \cite{sk,ad,mkz,mhd,bb}. Electrically charged thin-shell wormholes
in higher dimensional gravity  with a cosmological constant has been discussed in \cite{gas,fr}.
$d$-dimensional non-asymptotically flat thin-shell wormholes in Einstein-Yang-Mills-dilaton (EYMD)
gravity has been considered in \cite{shm} and its extension to Einstein-Yang-Mills-Gauss-Bonnet
was performed in \cite{shm,shmm}. Recently, an interesting well studied higher dimensional thin-shell wormholes have
been found in \cite{EE2012,GAS2010}. In this paper we are interested on studying $d$-dimensional thin-shell wormholes in a spherically symmetric spacetime surrounded by quintessence. Using the standard stability analysis, we will find the general stability conditions for that geometry and we will see that from our model, different other well-known thin-shell wormholes can be obtained with their respective stability conditions. 

This article is organized as follows: Sec. \ref{Sec1} establishes the basic notions for a $d$-dimensional spherically symmetric space-time surrounded by quintessence. In Sec. \ref{Sec2}, we construct thin-shell wormholes for this space-time by using the cut-and-paste technique. In addition, we directly derive the geodesic equation for a test particle which moves radially and initially is at rest. Sec. \ref{Sec4} is devoted to study the stability analysis using the standard linearised expansion method. Within this sections, we derive the general stability conditions for the thin-shell wormhole.
In Sec. \ref{Sec5} we demonstrate that from our general result, some interesting different particular cases studied before can be recovered. Additionally, we explore one new example and study the stability regions depending on the parameters. Finally, in Sec. \ref{Sec6} we conclude our main results.

\section{$d$-dimensional spherically symmetric  spacetime surrounded by quintessence}\label{Sec1}
Let us consider a $d$-dimensional static spherically symmetric metric surrounded by quintessence, which reads \cite{scbr}
\begin{equation}\label{1}
ds ^2 = -f(r) dt^2 + f(r)^{-1}dr^2+ r^2 d\Omega^{2}_{d-2}\,,
\end{equation}
where $\Omega^{2}_{d-2}$ represents the metric of the $(d-2)$ unit-sphere and the metric function is given by
\begin{equation}
 f(r)=\left[1-\frac{2M}{r^{d-3}}-\frac{C}{r^{(d-1)w_q+d-3}}\right]\,.\label{2}
\end{equation}

The space-time depends on the dimension $d$, the quintessence state parameter $w_q\leq 0$, the mass $M$ and a constant $C$. This geometry is the generalisation in higher dimensions of the well-known Kiselev solution in 4 dimensions (see \cite{Kiselev:2002dx,Fernando:2012ue,Zhang:2006ij,Younas:2015sva,Azreg-Ainou:2017obt} for some relevant studies related to it).
For instance, when $w_q = - 1$, the metric Eq. (\ref{1}) becomes
\begin{widetext}
\begin{equation}
ds ^2 = -\left[1-\frac{2M}{r^{d-3}}-Cr^2\right] dt^2 + \left[1-\frac{2M}{r^{d-3}}-Cr^2\right]^{-1}dr^2+ r^2 d\Omega^{2}_{d-2}\,,
\end{equation}
\end{widetext}
which reduces to the $d$-dimensional Schwarzschild de Sitter black hole where $C$ is the cosmological constant. Moreover, the
metric Eq. (\ref{1}) reduces to the $d$-dimensional Reissner-Nordstr$\text{$\ddot{o}$}$m black hole
if the quintessence state parameter takes the following form
\begin{equation}
w_q= \frac{d-3}{d-1}\,.
\end{equation}
In the following section, we will focus our study on the construction of thin-shell wormholes in this spacetime.\\

\section{construction of thin-shell wormhole and the gravitational field}\label{Sec2}
 We start the mathematical construction for our thin-shell
wormhole considering two identical copies of the vacuum solution
and removing from each copy the spacetime region given by
\begin{equation}
\Omega^{{\pm}}\equiv\lbrace r^{{\pm}}\leq a|\, a>r_{h} \rbrace\,,
\end{equation}
where $a$ is a radius constant grater than the event horizon $r_h$
to avoid the presence of horizon and singularities for the metric (\ref{1}). The removal of the regions from each spacetime gives two geodesically
incomplete manifolds, with the timelike hypersurfaces as boundaries given by
 \begin{equation}
\partial\Omega^{{\pm}}\equiv\lbrace r^{{\pm}}= a|\,a>r_{h} \rbrace\,.
\end{equation}
We obtain a geodesically complete manifold by identifying the
timelike hyperfurface $\partial\Omega^{{+}}=\partial\Omega^{{-}}$,
where two regions are connected by a wormhole. The identified region
$\partial\Omega$ is called the ``throat of the wormhole" where the exotic matter is
concentrated. The induced metric on the hypersurface $\partial\Omega$
takes the following form
\begin{equation}
ds ^2 = -d\tau^2 + a^{2}(\tau)d\Omega^2_{d-2}\,,
\end{equation}
where $\tau$ is the proper time along the hypersurface
$\partial\Omega$ and $a(\tau)$ defines the radius of the
throat as a function of the proper time. The surface stress
at the junction boundary are determined using the Darmois-Israel
formulation.

  The Lanczos equation gives the intrinsic surface stress-energy
tensor $S_{ij}$ which reads,
\begin{equation}
S^{i}_{j} = -\frac{1}{8\pi}\left(\kappa^i_j-\delta^i_j\kappa^m_m\right)\,,
\end{equation}
where the quantity $\kappa_{ij} = K^{+}_{ij}-K^{-}_{ij}$ represents the discontinuity in the
extrinsic curvature $K_{ij}^{\pm}$. Note that the symbol $-$ and $+$ corresponds to the interior
and exterior spacetime respectively. The second fundamental
form the extrinsic curvature can be defined as follows
\begin{equation}
K^{\pm}_{ij} = -\eta_{\nu}\left(\frac{\partial^2 x^{\nu}}{\partial\xi^i \partial\xi^j}+\Gamma^{\nu\pm}_{\alpha\beta}\frac{\partial x^{\alpha}}{\partial\xi^{i}}\frac{\partial x^{\beta}}{\partial\xi^{j}}\right)\,,\label{9}
\end{equation}
where $\eta_{\nu}$ represents the unit normal vector at the
junction and $\xi^i$ represents the intrinsic
co-ordinates. At the hypersurface $\partial\Omega$, the parametric
equation is given by $f\left(x^\mu(\xi^i)\right)=0$. By using this
equation, we derive the formula for the unit normal vector to the hypersurface $\partial\Omega$, which is given by
\begin{equation}
n_{\mu} = \pm\Bigl\lvert g^{\alpha\beta}\frac{\partial f}{\partial x^{\alpha}}\frac{\partial f}{\partial x^{\beta}}\Bigl\lvert ^{-\frac{1}{2}}\frac{\partial f}{\partial x^{\mu}}\,, \label{10}
\end{equation}
where the unitary condition $n_{\mu}n^{\mu}= +1$ holds and the discontinuity of the extrinsic
curvature $\kappa_{ij}$ can be written in a simplified form
due to spherical symmetry as
$\kappa^i_{j}= diag\left(\kappa^{\tau}_{\tau},\kappa^{\theta_1}_{\theta_1},......,\kappa^{\theta_{d-2}}_{\theta_{d-2}}\right)$.
Therefore, we can write the surface-energy tensor as
$S^i_j = diag\left(-\sigma, P,.....,P\right)$, where $\sigma$ is
the surface energy density and $P$ is the surface pressure.\\ Now, by using the Lanczos equation we find that
 \begin{equation}\label{11}
\sigma(a)=-\frac{(d-2)}{4\pi a}\sqrt{f(a)+\dot a^2}\,,
\end{equation}
and
 \begin{equation}\label{12}
P(a)=-\frac{d-3}{d-2}\sigma+\frac{ f^{\prime}(a)+\ddot a}{8\pi\sqrt{f(a)+\dot a^2}}\,,
\end{equation}
where primes and dots denote differentiation with respect to $a$ and $\tau$ respectively and the
function $f(a)$ is given in Eq. (\ref{2}). Here, the
energy density $\sigma$ and surface pressure $P$ obey
the following conservation equation
\begin{equation}
\frac{d}{d\tau}\left(\sigma a^{d-2}\right)+P\frac{d}{d\tau}\left( a^{d-2}\right)=0\,.\label{13}
\end{equation}
 For a static configuration of radius $a=a_0$, we have
$\dot{a} = 0 $ and $\ddot{a}= 0 $ so that, from  Eqs. (\ref{11}) and (\ref{12}) we directly find
\begin{equation}
\sigma(a_0)=-\frac{(d-2)}{4\pi a_0}\sqrt{f(a_0)}\,,\label{14}
\end{equation}
and
 \begin{equation}
P(a_0)=-\frac{d-3}{d-2}\sigma+\frac{ f^{\prime}(a_0)}{8\pi\sqrt{f(a_0)}}\,.\label{15}
\end{equation}
Let us now analyse the attraction and repulsive
nature of the wormhole on test particles. To do this, we calculate
the four-acceleration for the static wormhole $(\dot a =0)$, which
is written as
\begin{equation}
a^\mu = u^\mu_{\,\,;\nu} u^\nu\,,
\end{equation}
where the 4-velocity is $ u^\mu =\frac{d x^{\mu}}{d\tau}$= $(\frac{1}{\sqrt{f(r)}},0,0,.....,0)$. The only non-zero component of the acceleration is given by
\begin{equation}
a^r = \Gamma^r_{tt} \left(\frac{dt}{d\tau}\right)^2= \frac{M(d-3)}{r^{d-2}}+\frac{C[(d-1)w_q+d-3]}{2 r^{(d-1)w_q+d-2}}\,.
\end{equation}
Let us now consider a test particle which moves in radial direction and initially it is at rest. The equation of motion for this particle takes the following form
\begin{equation}
\frac{d^2r}{d\tau^2}= -\Gamma^r_{tt}\left(\frac{dt}{d\tau}
\right)^2 =-a^r\,,
\end{equation}
 which gives the geodesic equation if $a^{r}=0$. From here we can notice that
 the wormhole is attractive if $a^{r}> 0$ and repulsive if $a^{r}< 0$.

\section{ Linearized Stability analysis }\label{Sec4}

Equations of motion (\ref{11}) and (\ref{12}) can be rewritten in the following form
\begin{align}
\dot a^{2}-\frac{16\pi^2a^2}{(d-2)^2}\sigma^2&=-1+\frac{2M}{a^{d-3}}+\frac{C}{a^{(d-1)w_q+d-3}}\,,\label{19}\\
\dot \sigma&=-(d-2)\frac{\dot a}{a}\left(\sigma+P\right)\,.\label{20}
\end{align}
Now, if we integrate the energy conservation equation, Eq.~(\ref{13}), we get
\begin{equation}
\ln(a)=-\frac{1}{d-2}\int \frac{d\sigma}{\sigma+P(\sigma)}\,,
\end{equation}
which can be formally inverted to provide $\sigma=\sigma(a)$. To find the stability conditions for our configuration, we consider
linear perturbations around a static solution with radius $a_0$ \cite{po}. The surface energy density $\sigma(a_0)$ and the surface pressure $P(a_0)$ for the static solution are explicitly given in Eqs. (\ref{14}) and (\ref{15}) respectively. Now, the thin-shell equation of motion can be obtained by rewriting Eq. (\ref{19}) as
\begin{equation}
\dot a+V(a)=0\,,
\end{equation}
where the potential $V(a)$ is defined as
\begin{equation}
V(a)= f(a)-\frac{16\pi^2 a^2}{(d-2)^2}\sigma^2\,.\label{23}
\end{equation}
Since we are linearising around the static solution $a_0$,
we expand $V(a)$ around $a_0$ using Taylor series expansion up to
second order in powers of $(a-a_{0})$, which provides us
\begin{widetext}
\begin{eqnarray}
V(a) &=&  V(a_0) + V^\prime(a_0) ( a - a_0) +
\frac{1}{2} V^{\prime\prime}(a_0) ( a - a_0)^2  + \mathcal{O}\left[( a - a_0)^3\right]\,,\label{24}
\end{eqnarray}
\end{widetext}
where prime denotes derivatives with respect to $a$. The first order derivative of $V(a)$ is given by
\begin{equation}
V^{\prime}(a)= f^{\prime}(a)-\frac{32\pi^2 }{(d-2)^2}\left[\sigma+a\sigma^{\prime}\right]a\sigma\,.
\end{equation}

Now, if we use Eq. (\ref{20}), which is the conservation equation of the surface stress energy tensor, the above expression becomes
\begin{equation}
V^{\prime}(a)= f^{\prime}(a)+\frac{32\pi^2 }{(d-2)^2}a\left[(d-3)\sigma^2+(d-2)\sigma P \right]\,.\label{26}
\end{equation}

For the second derivative of the potential, we define a very useful
parameter $\eta(\sigma)=dp/d\sigma=P^{\prime}/\sigma^\prime$. Hence, the second derivative of the potential can be written as
\begin{widetext}
\begin{eqnarray}
V^{\prime\prime}(a)&=& f^{\prime\prime}(a)-\frac{32\pi^{2}}{(d-2)^2}
\Big\{
[(d-3)\sigma+(d-2)P]^{2} +(d-2)\left(d-3+(d-2)\eta\right) \sigma(\sigma+P)
\Big\}\,.
\end{eqnarray}
\end{widetext}
Since we are linearising around $a=a_0$, we can now go back to Eqs. (\ref{23}) and  (\ref{26}) to substitute for $a=a_0$ to find that $V(a_0)=0$ and $ V^\prime(a_0)=0$, respectively. Therefore, the potential $V(a)$ from Eq. (\ref{24}) is reduced to
\begin{equation}
V(a)= \frac{1}{2}V^{\prime\prime}(a_0)(a-a_0)^2+\mathcal{O}[(a-a_0)^3]\,,
\end{equation}
so that the equation of motion of the wormhole throat is given by
\begin{equation}
\dot a^2=-\frac{1}{2}V^{\prime\prime}(a_0)(a-a_0)^2+\mathcal{O}[(a-a_0)^3]\,.
\end{equation}

Thus, the wormhole is stable if and only if $V^{\prime\prime}(a_0)> 0$. Hence, $V(a_0)$ has a local minimum at $a_0$. To carry out this analysis, we can study which conditions we need for having stable wormholes. In our case, we find that this parameter needs to satisfy
\begin{equation}
 \eta_0 <\frac{a_0^{2}f_0^{\prime {2}}-2a_0^{2}f_0^{\prime \prime}f_0}{2(d-2)f_0(a_{0}f^{\prime}_0-2f_0)}-\frac{d-3}{d-2}\,,\label{30}
\end{equation}
where all the quantities with a suffix $0$ denote that they are evaluated at $a=a_{0}$. \\
Since we are interested in a more general scenario, we have to take into account the relation between the ADM mass $\mathcal{M}$ and the black hole mass parameter $M$ in the case of a $d$--dimensional spacetime. More specifically, in the case of a spherical geometry ($k=1$), we have the following relation
\begin{equation}
\mathcal{M}=\frac{16 \,\pi \, \Gamma (\frac{d-1}{2})}{(d-2) 2 \, \pi^{\frac{d-1}{2}}} M\,.
\end{equation}
Therefore, Eq. (\ref{30}) can also be written as
\begin{widetext}
\begin{equation}\label{32}
 \eta_0 <\frac{a_0^{2}f_0^{\prime {2}}-2a_0^{2}f_0^{\prime \prime}f_0}{2(d-2)f_0(-2f_{0}+\frac{\mathcal{M}(d-3)}{a_0^{d-3}}+\frac{C[(d-1)w_q+d-3]}{a_0^{(d-1)w_q+d-3}})}-\frac{d-3}{d-2},\,\,\,\text{if}\,\,\, -2f_{0}+\frac{\mathcal{M}(d-3)}{a_0^{d-3}}+\frac{C[(d-1)w_q+d-3]}{a_0^{(d-1)w_q+d-3}}>0
\end{equation}
and
\begin{equation}\label{33}
 \eta_0 > \frac{a_0^{2}f_0^{\prime {2}}-2a_0^{2}f_0^{\prime \prime}f_0}{2(d-2)f_0(-2f_{0}+\frac{\mathcal{M}(d-3)}{a_0^{d-3}}+\frac{C[(d-1)w_q+d-3]}{a_0^{(d-1)w_q+d-3}})}-\frac{d-3}{d-2}, \,\,\, \text{if}\,\,\, -2f_{0}+\frac{\mathcal{M}(d-3)}{a_0^{d-3}}+\frac{C[(d-1)w_q+d-3]}{a_0^{(d-1)w_q+d-3}}<0\,.
\end{equation}
\end{widetext}
These two above inequalities give us the condition where the wormhole is stable or not depending on all the parameters of the space-time.
Following Dias and Lemos approach \cite{gas}, we can write the general static metric solution for a $d$-dimensional solution with different $(d-2)$ geometric--topologies which are encoded by the geometric-topological
factor $k$. In that case, the function $f_0 \equiv f(a_0)$, given in Eq. (\ref{2}) evaluated at $a_0$ is given by
\begin{equation}
 f(a_0)=k-\frac{\mathcal{M}}{a_0^{d-3}}-\frac{C}{a_0^{(d-1)w_q+d-3}}\,,
\end{equation}
where we have three special cases: $k=1$ for spherical, $k=0$ for planar, and $k=-1$ for hyperbolic geometries. Morover for $f^{\prime}_0$ and $f^{\prime\prime}_0$ we have
\begin{equation}
f^{\prime}(a_0)=\frac{\mathcal{M}(d-3)}{a_0^{d-2}}+\frac{C[(d-1)w_q+d-3]}{a_0^{(d-1)w_q+d-2}}\,,
\end{equation}
and
\begin{widetext}
\begin{equation}
f^{\prime\prime}(a_0)=-\frac{\mathcal{M}(d-3)(d-2)}{a_0^{d-1}}-\frac{C[(d-1)w_q+d-3][(d-1)w_q+d-2]}{a_0^{(d-1)w_q+d-1}}\,.
\end{equation}
\end{widetext}

\section{ Special cases and new example} \label{Sec5}

  We now turn out our attention for some special cases which were
already studied by some authors. We will see that from  our general result, we can recover the same stability conditions  for a particular set of choices for the parameters. Additionally, a new interesting example is presented here.

\subsection{Poisson--Visser, $d=4$, $k=1$, $M\neq 0$, $C=0$ and $w_{q}=0$}

Let us start with the simplest wormhole stability solution, namely, the four-dimensional solution with spherical symmetry without charge and quintessence. In other words by setting  $d=4$, $k=1$, $M\neq 0$, $C=0$ and $w_{q}=0$ from Eq. (\ref{32}) and (\ref{33}) we find that the stability conditions become
\begin{equation}
 \eta_0 < \frac{-1+3M/a_{0}-3M^{2}/a_{0}^{2}}{2 \left(1-\frac{2M}{a_{0}}\right)\left(1-\frac{3M}{a_{0}}\right)}\,, \,\,\,\,\,\,\,\,\text{if} \,\,\,\,\, 1-\frac{3M}{a_{0}}>0
\end{equation}
and
\begin{equation}
 \eta_0 > \frac{-1+3M/a_{0}-3M^{2}/a_{0}^{2}}{2 \left(1-\frac{2M}{a_{0}}\right)\left(1-\frac{3M}{a_{0}}\right)}\,, \,\,\,\,\,\,\,\,\text{if} \,\,\,\,\, 1-\frac{3M}{a_{0}}<0\,.
\end{equation}

Note that in $d=4$, we have used the fact that $\mathcal{M}=2M$ since $\Gamma(3/2)=\sqrt{\pi}/2$. The above result first was found by Poisson and Visser in \cite{po}.

\subsection{Eiroa--Romero, $d=4$, $k=1$, $M\neq 0$, $C \neq 0$ and $w_{q}=1/3$}

Our solution can be extended to study the stability of four-dimensional spherical symmetry charged thin-shell wormhole, which corresponds to $d=4$ and $w_{q}=1/3$. From Eq. (\ref{32}) and (\ref{33}) it follows that this wormhole will be stable if
\begin{equation}
 \eta_0 < \frac{-1+3M/a_{0}-3M^{2}/a_{0}^{2}-MC/a_{0}^{3}}{2\left(1-\frac{2M}{a_{0}}-\frac{C}{a_{0}^{2}}\right)\left(1-\frac{3M}{a_{0}}-\frac{2C}{a_{0}^{2}}\right)}\,, 
\end{equation}
if
\begin{equation}\notag
1-\frac{3M}{a_{0}}-\frac{2C}{a_{0}^{2}}>0\,,
\end{equation}
and
\begin{equation}
 \eta_0 >  \frac{-1+3M/a_{0}-3M^{2}/a_{0}^{2}-MC/a_{0}^{3}}{2\left(1-\frac{2M}{a_{0}}-\frac{C}{a_{0}^{2}}\right)\left(1-\frac{3M}{a_{0}}-\frac{2C}{a_{0}^{2}}\right)}\,, 
\end{equation}
if 
\begin{equation}\notag
 1-\frac{3M}{a_{0}}-\frac{2C}{a_{0}^{2}}< 0\,.
 \end{equation}

Furthermore, if we replace $C=-Q^{2}$ in the above equations we recover the Reissner--Nordstrom TSW solution reported by  Eiroa and Romero in \cite{efe}.

\subsection{Lobo--Crawford, $d=4$, $k=1$, $M\neq 0$, $C \neq 0$ and $w_{q}=-1$}

We will now show that from Eqs (\ref{32}) and (\ref{33}) one can recover a spherically symmetric four-dimensional TSW solution with a cosmological constant. This case can be recoved by setting the quintessence parameter $w_{q}=-1$ and $d=4$. From Eqs (\ref{32}) and (\ref{33}) is not difficult to show that if
\begin{equation}
 \eta_0 < \frac{-1+3M/a_{0}-3M^{2}/a_{0}^{2}+3 C M a_{0}}{2\left(1-\frac{3M}{a_{0}}\right)\left(1-\frac{2 M}{a_{0}}-Ca_{0}^{2}\right)}\,, \,\,\,\text{if} \,\,\,a_{0}>3M
\end{equation}
and
\begin{equation}
 \eta_0 >  \frac{-1+3M/a_{0}-3M^{2}/a_{0}^{2}+3 C M a_{0}}{2\left(1-\frac{3M}{a_{0}}\right)\left(1-\frac{2 M}{a_{0}}-Ca_{0}^{2}\right)}\,, \,\,\,\text{if} \,\,\,a_{0}< 3M\,,
\end{equation}
the wormhole will be stable. Note that if we choose $C=\Lambda/3$, these inequalities are reduced to the solution found by Lobo and Crawford in \cite{fsn}.

\subsection{Rahaman--Kalam--Chakraborty, $d=d$, $k=1$, $M\neq 0$ and $w_{q}=(d-3)/(d-1)$ }

Further interesting generalizations can be recovered if we consider  a $d$-dimensional spacetime with spherical geometry with the quintessence parameter being $w_{q}=(d-3)/(d-1)$. Note that in this case we have a charged $d$-dimensional TSW with spherical geometry.  Here we also need to write the relation between the charge parameter $Q$ and the ADM charge $\mathcal{Q}$ which is given by
\begin{equation}
\mathcal{Q}^{2}=\frac{2}{(d-2)(d-3)}Q^{2}\,.
\end{equation}

Moreover, by setting $C=-\mathcal{Q}^{2}$ and by using the general conditions (\ref{32}) and (\ref{33}), it follows that the wormhole will be stable if
\begin{equation}
 \eta_0 <\frac{a_0^{2}f_0^{\prime {2}}-2a_0^{2}f_0^{\prime \prime}f_0}{2(d-2)f_0(-2+\frac{\mathcal{M}(d-1)}{a_0^{d-3}}-\frac{2\mathcal{Q}^{2}(d-2)}{a_0^{2(d-3)}})}-\frac{d-3}{d-2}\,, 
\end{equation}
if
\begin{equation}\notag
-2+\frac{\mathcal{M}(d-1)}{a_0^{d-3}}-\frac{2\mathcal{Q}^{2}(d-2)}{a_0^{2(d-3)}})>0\,,
\end{equation}
and
\begin{equation}
 \eta_0 >\frac{a_0^{2}f_0^{\prime {2}}-2a_0^{2}f_0^{\prime \prime}f_0}{2(d-2)f_0(-2+\frac{\mathcal{M}(d-1)}{a_0^{d-3}}-\frac{2\mathcal{Q}^{2}(d-2)}{a_0^{2(d-3)}})}-\frac{d-3}{d-2}\,, 
\end{equation}
if
\begin{equation}\notag
-2+\frac{\mathcal{M}(d-1)}{a_0^{d-3}}-\frac{2\mathcal{Q}^{2}(d-2)}{a_0^{2(d-3)}})<0\,.
\end{equation}

This solution in the literature was first studied by Rahaman--Kalam--Chakraborty \cite{fr}.

\subsection{Dias--Lemos, $d=d$, $k=1,0,-1$, $M\neq0 $ and $w_{q}=-1 $ }

Let us now recover the solution to a $d$-dimensional TSW with a cosmological constant and vanishing charge, i.e. $Q=0$. This result can be found from Eqs. (\ref{32}) and (\ref{33}) by setting the quintessence parameter $w_{q}=-1$. After some algebraic manipulation we find that the stability conditions become
\begin{equation}
 \eta_0 <\frac{a_0^{2}f_0^{\prime {2}}-2a_0^{2}f_0^{\prime \prime}f_0}{2(d-2)f_0(-2k+\frac{\mathcal{M}(d-1)}{a_0^{d-3}})}-\frac{d-3}{d-2}\,, 
\end{equation}
if 
\begin{equation}\notag
-2k+\frac{\mathcal{M}(d-1)}{a_0^{d-3}}>0\,,
\end{equation}
and
\begin{equation}
 \eta_0 >\frac{a_0^{2}f_0^{\prime {2}}-2a_0^{2}f_0^{\prime \prime}f_0}{2(d-2)f_0(-2k+\frac{\mathcal{M}(d-1)}{a_0^{d-3}})}-\frac{d-3}{d-2}\,, 
\end{equation}
if
\begin{equation}\notag
-2k+\frac{\mathcal{M}(d-1)}{a_0^{d-3}}<0\,.
\end{equation}
The $d$-dimensional TSW with a cosmological constant and three different geometries encoded by the parameter $k$ was recently investigated by Dias and Lemos. If we let $w_{q}=-1$ and $C=\Lambda /3$, the function $f_{0}=f(a_{0})$ can be written as
\begin{equation}
f(a_{0})=k-\frac{\mathcal{M}}{a_0^{d-3}}-\frac{\Lambda a_0^{2}} {3}\,,
\end{equation}
which corresponds to Dias and Lemos solution found recently in \cite{gas}.

\subsection{A new example, $d=d$, $k=1,0,-1$, $M\neq0 $ and $w_{q} \neq 0 $ }

Finally we can now consider a more general scenario, namely, a $d$-dimensional TSW wormhole surrounded by quintessence $w_{q} \neq 0 $. This case has not been considered yet in the literature. Note that since the energy density $\rho_{q}$ for quintessence should be positive and explicitly given as \cite{scbr}
\begin{equation}
\rho_{q}=-\frac{C w_q (d-1)(d-2)}{4 r^{(d-1)(w_{q}+1)}}\,,
\end{equation}
the quintessence parameter $w_q$, must be negative, i.e. $w_q \leq 0$. Hence, for this case, the stability conditions Eqs (\ref{32}) and (\ref{33}) related to three different geometries tells us that the TSW is stable when:
\begin{widetext}
\begin{equation}
 \eta_0 <\frac{a_0^{2}f_0^{\prime {2}}-2a_0^{2}f_0^{\prime \prime}f_0}{2(d-2)f_0(-2k+\frac{\mathcal{M}(d-1)}{a_0^{d-3}}+\frac{C(d-1)(w_q+1)}{a_0^{(d-1)w_q+d-3}})}-\frac{d-3}{d-2},\,\,\,\,\,\,\,\,\text{if} \,\,\,\,\,-2k+\frac{\mathcal{M}(d-1)}{a_0^{d-3}}+\frac{C(d-1)(w_q+1)}{a_0^{(d-1)w_q+d-3}}>0,
\end{equation}
and
\begin{equation}
 \eta_0 > \frac{a_0^{2}f_0^{\prime {2}}-2a_0^{2}f_0^{\prime \prime}f_0}{2(d-2)f_0(-2k+\frac{\mathcal{M}(d-1)}{a_0^{d-3}}+\frac{C(d-1)(w_q+1)}{a_0^{(d-1)w_q+d-3}})}-\frac{d-3}{d-2},\,\,\,\,\,\,\,\,\text{if} \,\,\,\,\,-2k+\frac{\mathcal{M}(d-1)}{a_0^{d-3}}+\frac{C(d-1)(w_q+1)}{a_0^{(d-1)w_q+d-3}}<0\,.
\end{equation}
\end{widetext}
For a useful information on the wormhole stability in different geometries and different dimensions we show graphically the wormhole stability in terms of the parameter $\eta_{0}$ and $a_0$, for different values of $d$, $\mathcal{M}$, $C$ and $w_q$.

In Fig. 1 we show the stability region for the spherical geometry, i.e. $k = 1$. As we can see, in the first and second plot there are two interesting intervals that are worth of mentioning. The first interval is between the left and the right asymptote where the region of stability is located above the curve. The second interval is on the right to the right asymptote with the region of stability being located below the curve. In the third plot, there is only one interval worth of mentioning with the region of stability below the curve.

In Fig. 2 we show graphically the stability region for the planar geometry, i.e. $k = 0$. There are two important intervals for the first and second plot and only one interval for the third plot. The corresponding region of stability in the first case is above the curve, while in the later case the region of stability is below the curve.

Finally, in Fig. 3 the stability region for the hyperbolic geometry ($k = -1$) is depicted. From the first and second plot we can easily observe that the regions of stability are above (below) the curve in the first (second) interval. Last but not least, we are left with only one interesting interval with the region of stability below the curve in the last plot. From those examples if follows that,  the stability domain of the HDTSW increases if we increase the number of dimensions.

\begin{figure}[H]
\includegraphics[width=0.30\textwidth]{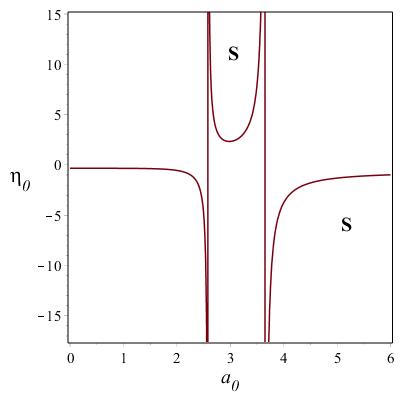}\label{fig1} %
\includegraphics[width=0.30\textwidth]{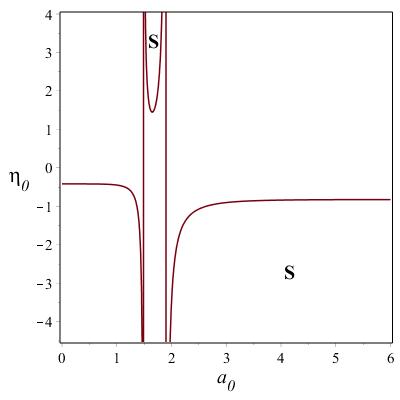}\label{fig2} %
\includegraphics[width=0.30\textwidth]{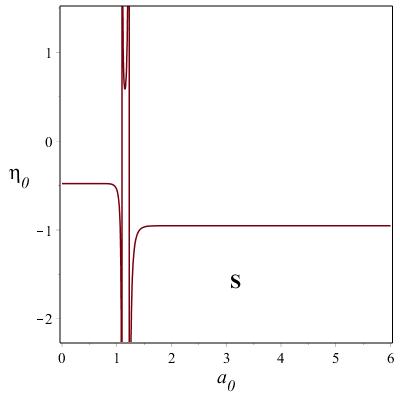}\label{fig3} %
\caption{\small Stability regions of HDTSW for $k=1$ and $M=10$ in all cases. In the first case we have chosen $d=5$, $C=-0.5$, and $w_{q}=-0.5$. In the second case $d=8$, $C=-0.5$, and $w_{q}=-0.6$ while in the last plot we have chosen $d=24$, $C=-0.5$, and $w_{q}=-0.9$. The stability region is denoted by $S$.}
\end{figure}

\begin{figure}[H]
\includegraphics[width=0.30\textwidth]{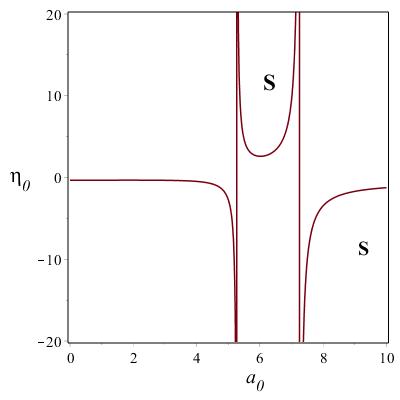}
\includegraphics[width=0.30\textwidth]{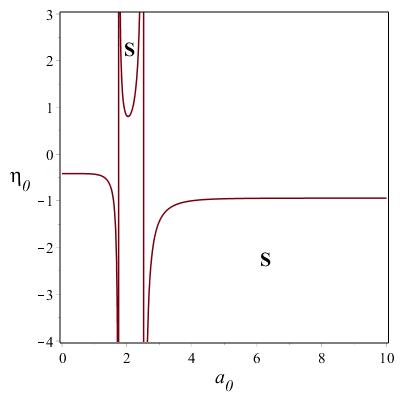}
\includegraphics[width=0.30\textwidth]{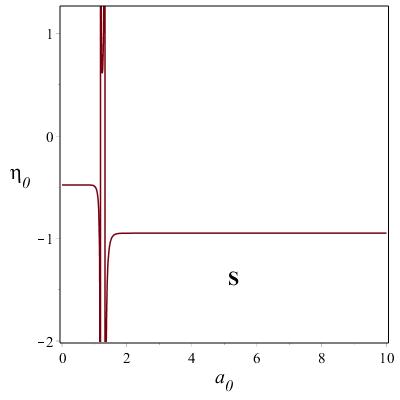}
\caption{\small Stability regions of HDTSW for $k=0$ and $M=10$ in all cases. In the first case we have chosen $d=5$, $C=-0.7$, and $w_{q}=-0.4$. In the second case $d=8$, $C=-0.3$, and $w_{q}=-0.9$ while in the last plot we have chosen $d=24$, $C=-0.5$, and $w_{q}=-0.9$. The stability region is denoted by $S$.}
\end{figure}

\begin{figure}[h!]
\includegraphics[width=0.30\textwidth]{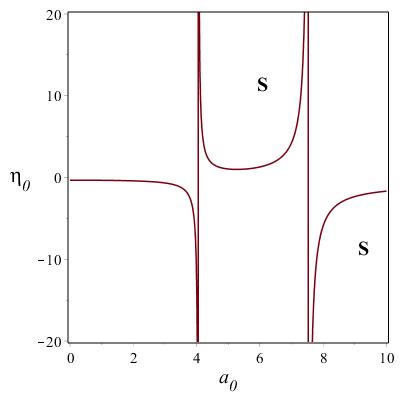}
\includegraphics[width=0.30\textwidth]{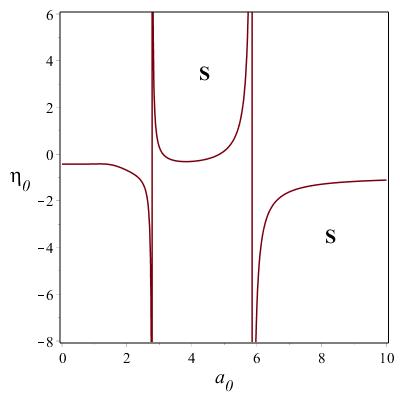}
\includegraphics[width=0.30\textwidth]{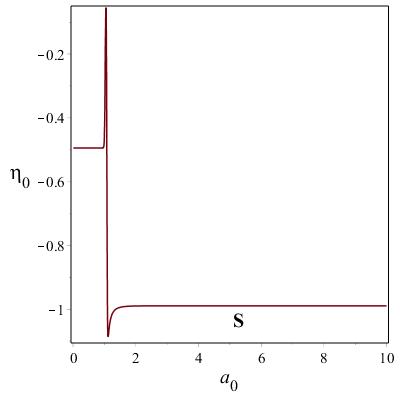}
\caption{\small Stability regions of HDTSW for $k=-1$ and $M=10$ in all cases. In the first plot we have chosen $d=5$, $C=-0.3$, and $w_{q}=-0.8$. In the second plot $d=9$, $C=-0.3$, and $w_{q}=-0.9$ while in the third plot we have chosen $d=100$, $C=-0.3$, and $w_{q}=-0.9$. The stability region is denoted by $S$. }
\end{figure}

\section{CONCLUSIONS} \label{Sec6}

In this work we  have used the cut-and-paste method and the Darmois-Israel formalism to construct a $d$-dimensional thin-shell wormhole surrounded by quintessence. Taking into account the standard junction conditions we have investigated a more general class of stability conditions under radial perturbations preserving the spherically symmetry in three different geometries and explore the stability regions in different dimensions. In particular we investigated the wormhole stability regions for a spherical geometry, planar geometry, and hyperbolic geometry. Finally we discuss the stability regions in different dimensions $d$ and investigate the effects of different values of the parameters $w_{q}\leq 0$, $\mathcal{M}$, and $C$ for the stability of the wormhole. For all three different geometries it is shown that by increasing the number of dimensions, increases the stability domain for obtaining stable $d$--dimensional wormholes surrounded by quintessence. In particular,  for a given state parameter $w_{q}\leq 0$ of the quintessence, we observe from Figures 1-3 that as the number of dimensions $d$ increases, the gap between the asymptotes get shorter. The other peculiarity of our $d$-dimensional wormhole solution, is the assumption that our $d$-dimensional wormhole is surrounded by the quintessence matter which is not only on the brane but also in the bulk. This means that, essentially, we have found a more general wormhole solution and studied the stability domain of our wormhole with the role played by the quintessence which is also known as one of the possible candidates of  dark energy in cosmology. 

\bigskip
\begin{flushleft}
	\textbf{Acknowledgments}
\end{flushleft}
S.B. is supported by the Comisi{\'o}n Nacional de Investigaci{\'o}n
Cient{\'{\i}}fica y Tecnol{\'o}gica (Becas Chile Grant
No.~72150066).

\end{document}